\documentclass[aip,graphicx,jap]{revtex4-1}

\usepackage{amssymb}
\usepackage{graphicx}

\def\nevreal{(\omega_0^2 - \omega^2)^2 + \gamma_2^2 \omega^2}

\def\SBox{(\Omega/\Box)^{-1}}

\draft 

\begin{document}


\title{A general figure of merit for thick and thin transparent conductive carbon nanotube coatings}



\author{\'A. Pekker}

\author{K. Kamar\'as}
\email[]{kamaras@szfki.hu}
\affiliation{Research Institute for Solid State Physics and Optics, Hungarian Academy of Sciences, P.O. Box 49, Budapest, Hungary H-1525}

\date{\today}

\begin{abstract}
We suggest a wavelength-dependent figure of merit for transparent conducting nanotube networks, composed of the sheet resistance and the optical density. We argue that this would be more useful than other suggestions prevailing in the literature, because it relies on more realistic assumptions regarding the optical parameters of real nanotubes: it takes into account the fact that the dc resistivity depends on the concentration of free carriers, while the visible absorption is caused by bound carriers. Based on sheet resistance measurements and wide-range transmission spectra, we compare several commercial nanotube types and find correlation between metal enrichment and figure of merit. A simple graphical approach is suggested to determine if the required optical and transport properties can be achieved by varying the thickness of the nanotube layer or a more aggressive
treatment is needed. The procedure can be extended to oxide coatings as well.
\end{abstract}


\maketitle

\section{Introduction}
\label{Intro}
One of the most successful applications of carbon nanotubes is that of transparent conductive coatings. These systems were extensively investigated\cite{gruner06} and several manufacturing techniques have been suggested. \cite{lee04,saran04,barnes07,hu04,zhang06a,hu07,li07,ma07,parekh07}
Of special interest are thin films from density-gradient separated samples, which form colored semitransparent coatings of selected chiralities.\cite{green08,yanagi08}

With the development of these products came naturally the demand for unique characterization of film quality. Two parameters have to be balanced in a transparent conductive layer: the transmission in a chosen spectral region (for the most important application, in
solar cells, this is preferably the spectrum of the solar radiation) and the dc conductivity (or sheet resistance). The film is considered higher quality if the transparency is high and the sheet resistance low; both quantities depend on the optical and electrical properties and the thickness \emph{d} of the film. Of course, for the film to be conductive at all, \emph{d} has to reach a certain minimum value; this minimum thickness gives the percolation threshold which has been reported for several kinds of nanotubes.\cite{bekyarova05,hu04} Above this threshold, many parameters influence the optimization, e.g. mechanical stability; nevertheless, sheet resistance and transmission at some chosen wavelength remain the principal ingredients which are desirable to be combined into a single figure of merit.

Two approaches can be used for such a purpose: either a completely pragmatic one, just bearing in mind that the characteristic quantity should reflect the feasibility for practical applications, or one which relies on more fundamental optical relationships and combines them using realistic approximations. The simplest example for the first approach was that of Fraser and Cook,\cite{fraser72} who used the ratio of transmission at a given wavelength and the sheet resistance to characterize their ITO films. Although this figure was useful for ITO coatings in the thickness range 0.2 to 2 $\mu$m, it was later shown by Haacke\cite{haacke76} that in general cases it reaches the optimum value at a thickness where the transmission is 0.37, clearly a bad choice for transparent conductive coatings. Haacke suggested to use the figure of merit $\Phi_{TC} = T^{10}/R_\Box $, which shows maximum at T=0.9. He showed that in certain, well-defined cases $\Phi_{TC}$ can be derived from the fundamental materials parameters, but in any general case a higher $\Phi_{TC}$ means better quality.  The idea behind the procedure is to emphasize the importance of optical transparency, since sheet resistance can vary in a much wider range for the film to be a good transparent conducting material. The choice of the exponent reflected the limits of the transparency of available coatings more than thirty years ago; since then, new materials appeared exhibiting higher than 90 per cent transmission for which the Haacke figure of merit breaks down.

A significant improvement, using the second approach, was the suggestion of Gordon\cite{gordon00} to use the ratio $\sigma/\alpha$, of the inverse of the sheet resistance and the visible absorption coefficient, the latter calculated from the total visible transmission and corrected for reflectance. He applied this quantity succesfully to compare inorganic oxide coatings.

The most popular method, proposed specifically for nanotubes, based on a more fundamental description of optical properties of solids, is that of Hu, Hecht and Gr\"uner.\cite{hu04} They start from a simplified version of the formula introduced by Tinkham:\cite{glover57,tinkham96}
\begin{equation}
T(\omega) = \frac{1}{(1+\frac{2\pi}{c}\sigma_1(\omega)d)^2+ (\frac{2\pi}{c}\sigma_2(\omega)d)^2},
\label{eq:tinkham}
\end{equation}
neglect the imaginary part of the conductivity $\sigma_2$, and substitute the thickness from the expression connecting the dc sheet resistance and the dc conductivity $d=\frac{1}{R_\Box\sigma_1(0)}$:
\begin{equation}
\label{eq:dresselgruner}
T(\omega) = \frac{1}{(1+\frac{2\pi}{c}\sigma_1(\omega)d)^2}=\frac{1}{(1+\frac{2\pi}{cR_\Box(0)}\frac{\sigma_1(\omega)}{\sigma_1(0)})^2}.
\end{equation}

(Throughout this paper, we do not formulate the optical functions vs. wavelength, as is more usual in applications in the visible range, but vs. frequency; in the latter representation, the transition from ac to dc is more straightforward.)

\begin{figure}[h]
\centering

\includegraphics[width=7cm]{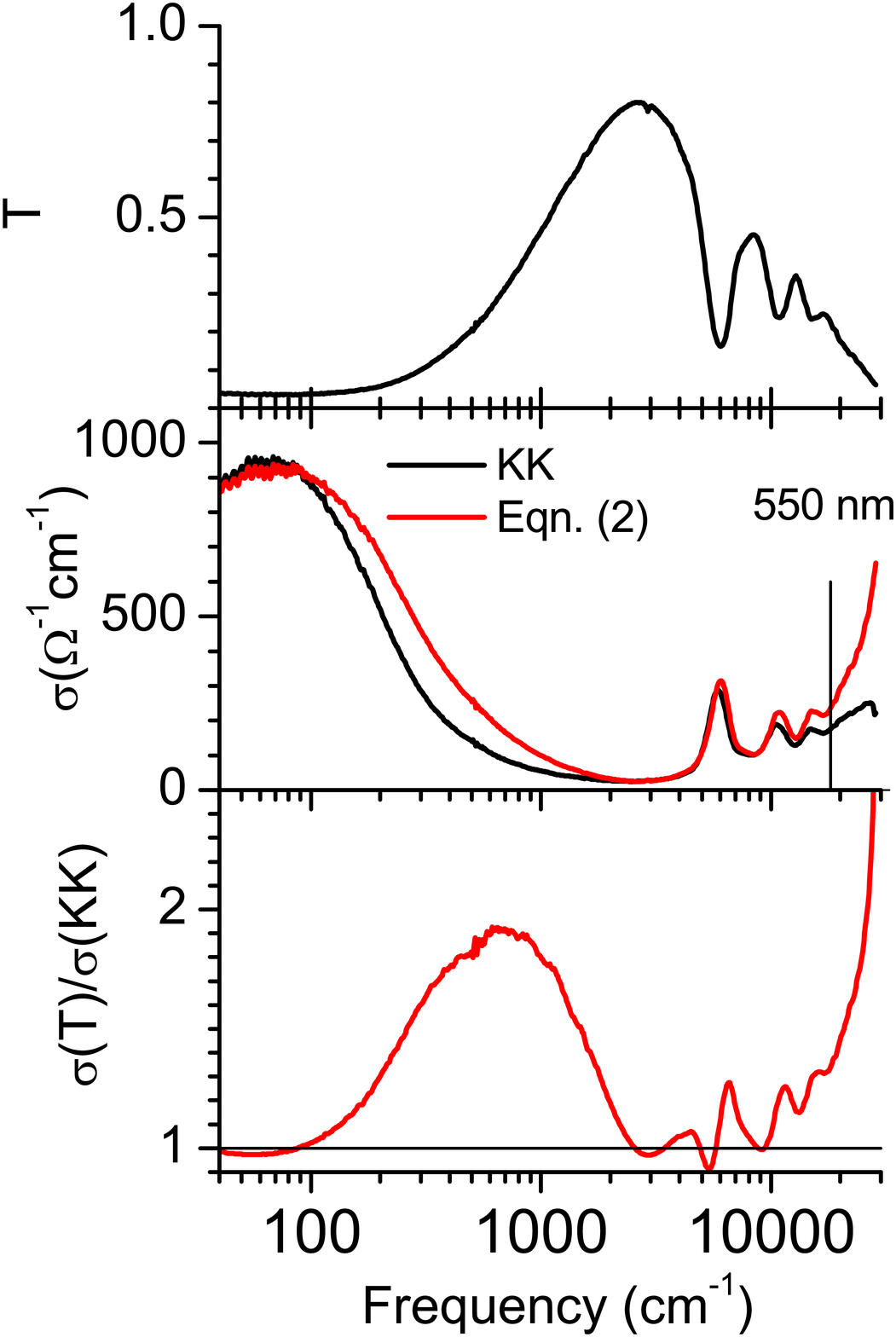}

\caption{Top: Transmission vs. frequency of a 250 nm thick  nanotube film synthesized by laser ablation (from Ref. \onlinecite{borondics06}). Middle: Optical conductivity calculated from the top curve by Kramers-Kronig transformation (black) and by the formula given in Ref. \onlinecite{hu04}. Bottom: the ratio of the two curves in the middle panel.}
\label{fig:sigma}
\end{figure}

Equation \ref{eq:tinkham} is intended to be used for thin metallic films in the microwave and far infrared range, where $\sigma_1 \gg \sigma_2$, and the optical functions are independent of frequency, i.e. $\frac{\sigma_1(\omega)}{\sigma_1(0)}=1$. (The validity has been tested in Ref. \onlinecite{glover57} for wavelengths above 100 $\mu$m, corresponding to frequencies below 100 cm$^{-1}$.) At higher frequencies, where the conductivity is complex and frequency-dependent, the simple relationship between transmission and conductivity is not valid anymore; the conductivity has to be determined from wide-range transmission data using the Airy formula\cite{dresselgruner02} and Kramers-Kronig transformation. The result of the latter procedure is illustrated in  Fig. \ref{fig:sigma} for a 250 nm freestanding nanotube network synthesized by laser ablation.\cite{borondics06}  The optical properties of this network could be fitted in a quite satisfactory way with a Drude-Lorentz model, where the Drude part represents the contribution from the free carriers in metallic tubes and the Lorentzian curves those of the interband transitions (of bound carriers in both metallic and semiconducting tubes) and the $\pi \rightarrow \pi^*$ excitation. For the nanotube network illustrated in Fig. \ref{fig:sigma}, Eqn. \ref{eq:dresselgruner}  overestimates the exactly calculated value by up to a factor of 2 and shows a marked frequency dependence. Given that the spectra of nanotubes can vary widely depending on preparation conditions,\cite{kamaras08} it is obvious that the $\sigma_1(\omega)/\sigma_1(0)$ value cannot be chosen as a frequency-independent parameter. If we consider it frequency-dependent, however, then the conditions for the validity of Eqn. \ref{eq:dresselgruner} are not fulfilled.

This choice is extremely dangerous when used for comparing networks of different composition whose spectrum can differ fundamentally. Indeed, when evaluating metal-enriched nanotubes by this method\cite{green08} it was found that the estimated dc conductivity enhancement depends on the wavelength where the transmission is measured. Such results are clearly unphysical, since the electrical conductivity is a fundamental property of the material and cannot depend on measurement conditions. The evaluation method is now spreading into the literature and used for thick and thin films, without consideration for the boundary conditions. In some publications, it is thought that the formula can be used so long as the film is thinner than the measuring wavelength (this condition is only valid together with the $\sigma_1 \gg \sigma_2$ one), in others that it is valid for high transmission values only\cite{liu08} (although from Fig. \ref{fig:sigma} it is obvious that at low frequencies, where the transmission is very low, the formula works very well).

\begin{figure}[h]
\centering

\includegraphics[width=7cm]{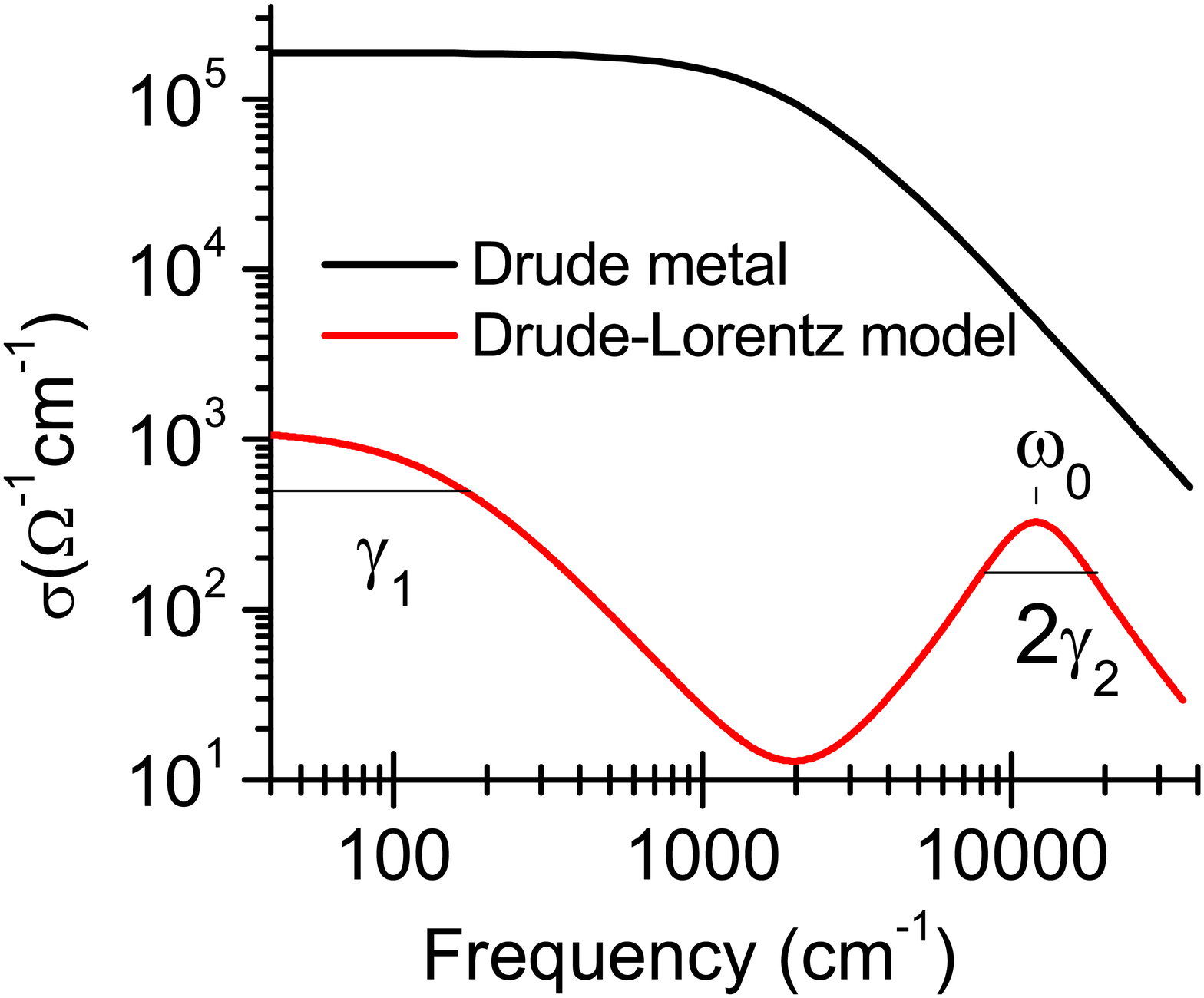}

\caption{Optical conductivity of an ideal Drude metal ($\omega_p=15000\ cm^{-1}, \gamma=2000\ cm^{-1}$) and a two-component model system ($\omega_{p1}=3200\ cm^{-1}, \gamma_1=150\ cm^{-1}, \omega_{p2}=14000\ cm^{-1}, \omega_0=12000\ cm^{-1}, \gamma_2=10000\ cm^{-1}$).}
\label{fig:model}       
\end{figure}

Figure \ref{fig:sigma} also shows that from the point of view of optical excitations, a nanotube network is not a metal at 550 nm, where visible transmission is measured most often. Assuming a simple Drude model, the difference between ordinary metals and metallic nanotubes is that in the latter, both the plasma frequency and the relaxation rate are much lower (Fig \ref{fig:model}). This means that while the dc conductivity reflects the properties of free electrons, the absorption in the NIR/visible is caused by bound electrons: the interband transitions in both semiconducting and metallic nanotubes (these include but are not restricted to the ones causing the far-infrared absorption and conductivity) and the $\pi \rightarrow \pi^*$ excitation. Thus, the far infrared and visible processes are decoupled. Equation \ref{eq:dresselgruner} describes the far-infrared part, the dynamics of the free carriers; above $\approx$ 2000 cm$^{-1}$ (below 5000 nm), $\sigma_1(\omega) = 0$ for these carriers. (The case is similar in ITO and other conducting oxides, where the conductivity is caused by free carriers with a low relaxation rate: as stated by Gordon,\cite{gordon00} above the plasma frequency these materials behave as transparent dielectrics.)

A possible way out of the vicious circle described above is to use a figure of merit which can be derived from transmittance at any frequency and dc sheet resistance but does not suffer from the limitations mentioned. We suggest a modified version of the value introduced by Gordon\cite{gordon00} for inorganic oxide layers: the ratio of the optical density \emph{at a given frequency} $-log\ T(\omega)$ and the dc sheet conductance $S_\Box = \frac{1}{R_\Box}$. (The original suggestion\cite{gordon00} includes the total visible transmission and reflection values, which are widely used in optical engineering but not in the case of nanotube films.) We build on the simple fact that both quantities show a linear dependence on film thickness, with certain restrictions. For the optical density, these are fulfilled in the visible and at lower wavelengths, where reflectivity corrections do not influence the measured optical density,\cite{pekker06} which is also the desired range for most applications. In the case of inorganic oxides,\cite{gordon00} where even the thinnest layers provide uniform coverage, measurement of one point is enough to determine the ratio $\sigma/\alpha$. In the special case of nanotubes, where a distinct percolation threshold exists, this figure of merit can be accurately determined only by measuring a series of samples with varying thickness; however, since the relationship is linear, 3-4 points suffice to arrive at a reliable value.

\section{Model and figure of merit}

We start with a system described by one Drude and one Lorentz oscillator (Fig. \ref{fig:model}). This is a simplified model for a nanotube ensemble where the free carriers in the metallic tubes show Drude behavior, confined to the infrared (plasma frequency $\approx$ 2000 cm$^{-1}$), and the higher-frequency transitions are caused by bound carriers in both metallic and semiconducting tubes. The parameters used here are similar to those found in Ref. \onlinecite{borondics06}: the Drude curve represents free carriers in metallic tubes and the first Lorentz oscillator bound carriers in semiconducting tubes. (We neglect the higher-lying interband transitions from both semiconducting and metallic tubes, and the $\pi \rightarrow \pi^*$ excitations of the whole $\pi$-electron system.) For comparison, we also show the optical conductivity of an ideal Drude metal with parameters close to that of aluminum. The conductivity of the two-component system is
\begin{equation}
\label{eq:optcond}
\sigma_1 = {{N_1e^2} \over {mV}}{{\gamma_1} \over {(\omega^2 + \gamma_1^2)}}  +  {{N_2e^2} \over {mV}}
{{\gamma_2 \omega^2} \over {\nevreal}}
\end{equation}
where V is the volume of the system, $N_1$, $m$ and $\gamma_1$ the number of free carriers, the electron mass (we assume it to be equal to the effective mass of the carriers) and the width of the free-carrier conductivity (the relaxation rate), respectively, and $N_2$ and $\gamma_2$ the same quantities for the bound carriers (these can be electrons in the Van Hove singularities or the $\pi \rightarrow \pi^*$ transition). There is no \emph{a priori} reason in a nanotube ensemble why $N_1$ and $N_2$ should be related in any way, since the two transitions are caused largely by different nanotubes in the network. The conditions of Eqn. \ref{eq:tinkham} are fulfilled for frequencies $\omega < \gamma_1$.\cite{glover57,tinkham96}

Assuming that the dc conductivity can be given by the zero limit of the optical conductivity (the same as in the metallic limit), we obtain for the sheet conductance $S_\Box$
\begin{equation}
\label{eq:dccond}
S_\Box = \sigma_1(0)d = {{N_1e^2} \over {mV\gamma_1}}d.
\end{equation}

Alternatively, we can express the conductivity by way of the carrier mobility of the free carriers:
\begin{equation}
\label{eq:mobil}
\sigma_1(0) = {{N_1e^2} \over {mV\gamma_1}} = {\frac{N_1}{V}}e\mu,
\end{equation}
where
\begin{equation}
\label{eq:mobildef}
\mu=\frac{e}{m\gamma_1}.
\end{equation}
Note that only the free carriers (type 1) contribute to the dc conductivity and the mobility.

In the frequency range of the Lorentzian part, the reflectivity is low enough to be neglected and the optical density $-log\ T$ obeys Beer's law:\cite{pekker06}
\begin{equation}
\label{eq:optdens}
-log\ T = \epsilon(\omega) \frac{N_2}{V}d
\end{equation}
with the factor $\epsilon(\omega)$ usually termed extinction coefficient in analytical chemistry (not to be confused with the imaginary part of the refractive index \emph{k} for which the same notation is used).

From these two equations, it follows that $S_\Box$ will depend linearly on $-log\ T$, with the slope
\begin{equation}
\label{eq:slope}
M(\omega) = \frac{S_\Box}{-log\ T(\omega)} = \frac{N_1}{N_2}\frac{1}{\epsilon(\omega)}\frac{e^2}{m\gamma_1}.
\end{equation}

The slope, in this ideal case, thus bears a physical meaning: the ratio between the number of free carriers causing the dc conductivity and the number of bound carriers causing the high-frequency absorption. It is a frequency- (or wavelength)-dependent quantity where the wavelength dependence can be predicted from the transmission spectrum through $\epsilon(\omega)$. It can also be generalized to cases where more than one Lorentzian is giving contributions at a given wavelength, provided the reflectance is not too high so that the optical density depends linearly on the amount of nanotubes in the sample. This can be the case for pristine networks but also for composite materials where the optical density is linear in concentration.\cite{wang08}

Even in less-than-ideal cases where the dc conductivity is determined by contacts and thus does not correspond to the zero-frequency value of the Drude contribution, the figure of merit still can be used for comparison of different materials. Instead of  Eqn. \ref{eq:slope} the following applies:
\begin{equation}
\label{eq:percslope}
M(\omega) = \frac{\sigma_{dc}d}{-log\ T(\omega)} = \frac{V}{N_2}\frac{1}{\epsilon(\omega)}\sigma_{dc}
\end{equation}
Equation \ref{eq:percslope} contains $\sigma_{dc}$ as a constant typical of the given network. In this case, of course, physical quantities like free carrier concentration cannot be derived from the measured data.

\begin{figure}[h]
\centering

\includegraphics[width=7cm]{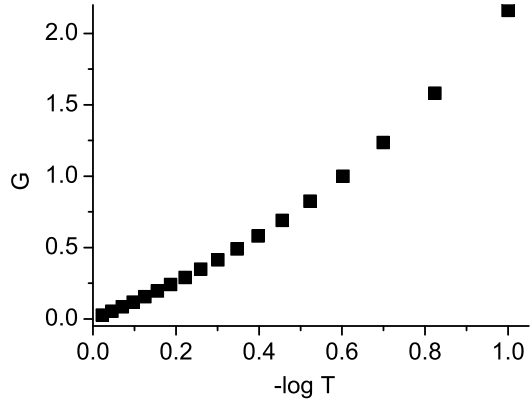}

\caption{Optical density vs. the function $G=((1-\sqrt{T})/\sqrt{T}$.}
\label{fig:compare}       
\end{figure}

The meaning of Eqn. \ref{eq:dresselgruner} can also be generalized and used as a simple mathematical algorithm to compare different transparent conductors. The detailed recipe was given by Liu et al.,\cite{liu08} who rearranged Eqn. \ref{eq:dresselgruner} in order to obtain a characteristic parameter from a linear least-square fit. Using our notation, the rearranged form is:
\begin{equation}
\label{eq:Gdef}
{S_\Box} = \frac{c}{2\pi\beta}G = \frac{c}{2\pi\beta}\frac{1-\sqrt{T}}{\sqrt{T}}
\end{equation}

In this approach, the quantity derived from the transmittance is $G = (1-\sqrt{T})/\sqrt{T}$, while in the one proposed by us, it is the optical density based on Beer's law.
Figure \ref{fig:compare} shows that the two functions are monotonous for any transmission value and linear for about T values larger than 40 per cent. Thus the relative characterization of different networks is not influenced by using a different figure of merit. One has to bear in mind, however, that for thicker networks the value $\beta$ becomes thickness-dependent, as already noted by Liu et al.\cite{liu08}, and the boundary conditions of Eqn. \ref{eq:dresselgruner} have to be fulfilled if one tries to derive dc conductivity values.

\begin{figure}[h]
\centering

\includegraphics[width=7cm]{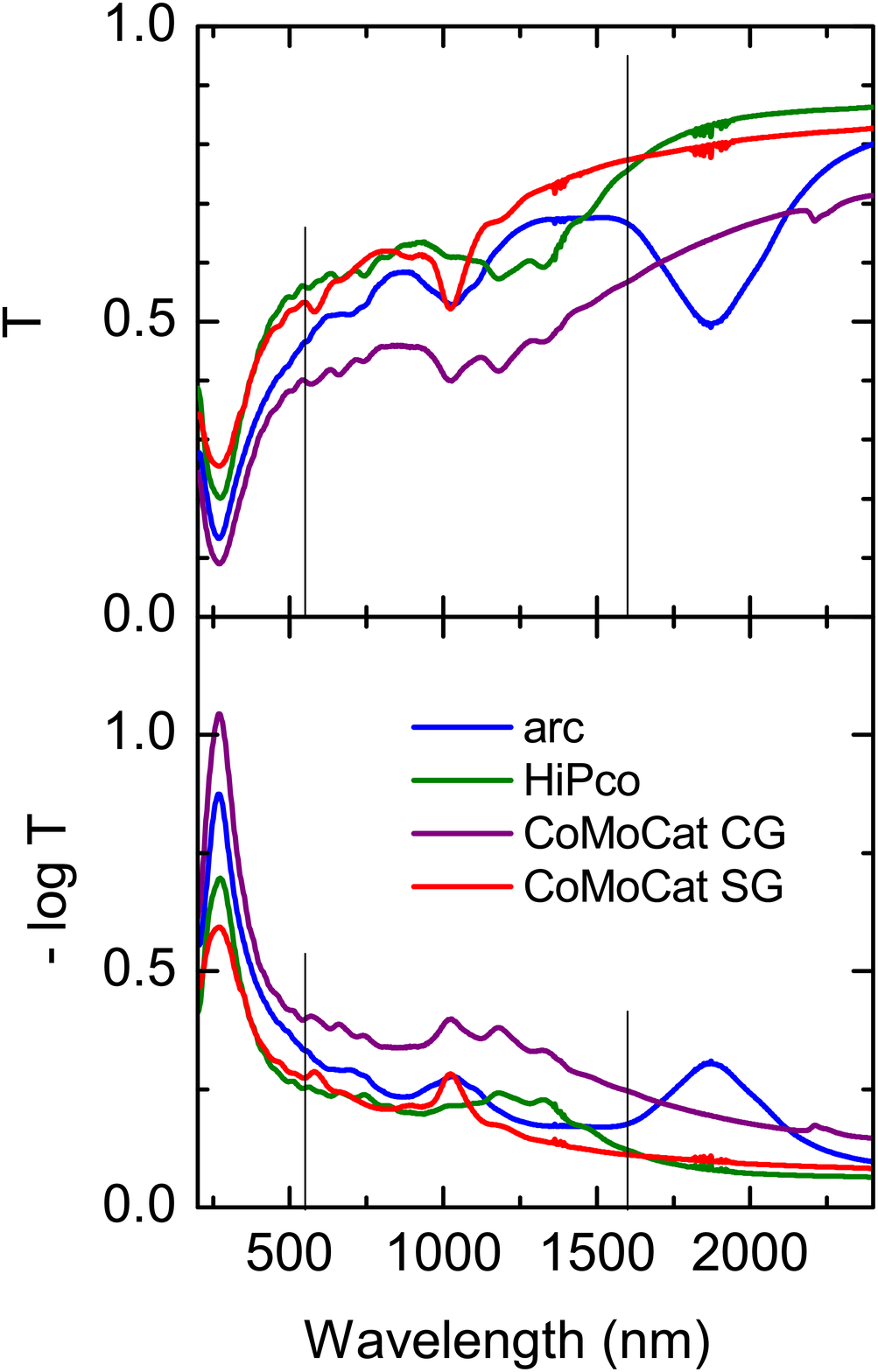}

\caption{Frequency-dependent transmission (top panel) and optical density (bottom panel) for our thinnest networks in the NIR/VIS range.}
\label{fig:Tspectra}       
\end{figure}

\begin{figure*}[t]
\centering

\includegraphics[width=12cm]{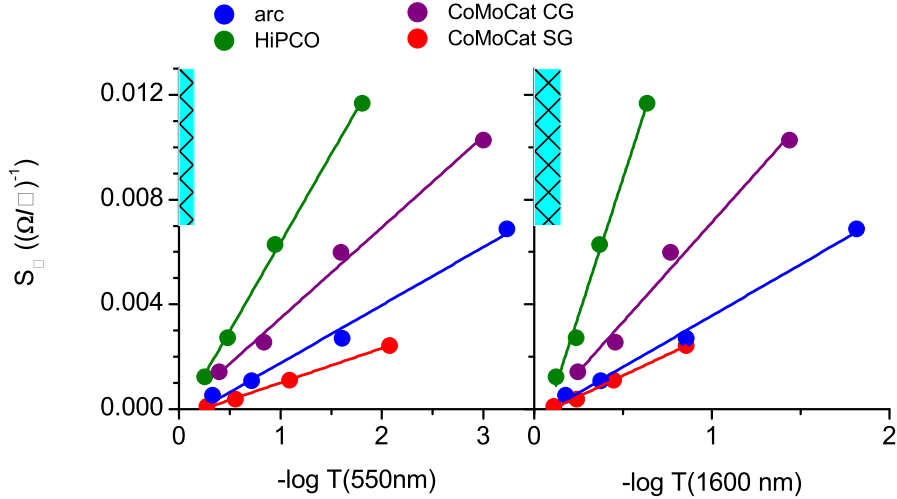}

\caption{Dc sheet conductance $S_\Box$ vs. (left panel) ${-log T}$(550 nm) and (right panel) ${-log T}$(1600 nm) for several unsorted thick nanotube networks). The lines are results of a linear fit according to Eqn. \ref{eq:slope}. The shaded area represents the desired region of technological applications.}
\label{fig:ours}       
\end{figure*}

The $M(\omega)$ value defines a figure of merit at one specific wavelength and not the overall transparency in one given spectral range; wavelength-dependent curves can be readily constructed from the transmission spectrum of the sample and will scale with the inverse of the optical density ${-log T}$.

We know of two papers in the recent literature suggesting similar figures of merit based on optical density and sheet resistance. Eigler\cite{eigler09} introduces a number which only differs from the one presented here in a scaling factor, equal to the theoretical extinction coefficient of one graphene layer, 301655 cm$^{-1}$. The argument for this choice is that the graphene materials constants, which can be derived from fundamental quantities, can serve as a general basis for any carbon-based system. However, we think that even among organic materials, variations in optical properties are too large for the use of a single reference material to be justified, not to speak of inorganic coatings like ITO.

Dan, Irvin and Pasquali\cite{dan09} derive a figure of merit which is the inverse of the one suggested here (apart from a different base of logarithm; our reasoning behind the choice of $log T$ instead of $ln T$ is simply that spectrometer manufacturers opted for the former in their software and therefore the optical density used in our approach is the same as the numerical value measured with any commercial instrument in "absorbance" mode).

We see the advantage of the suggested M($\omega$) values over the two presented above in that they allow a more straightforward graphical comparison of transparent conductors. In the next section we illustrate this advantage for several applications.

\section{Experimental test of the model}

To demonstrate the effect of sample composition on the quality of transparent conductive coatings, we prepared four series of thick networks from various commercially available types of single-walled nanotubes on quartz substrates. We chose nanotubes prepared by arc discharge (Carbon Solutions P2), HiPCo and two types of CoMoCat (Southwest Nanotechnologies) samples: commercial grade with a wider distribution of tubes and special grade enriched in (6,5) and (7,5) tubes. We prepared several samples by vacuum filtration.\cite{wu04}
Usually 5mg of nanotube was suspended in 500 ml distilled water mixed with 10 ml Triton-X surfactant (Triton-X100-Aldrich). A bath-type sonicator was used to homogenize the
suspension (sonication time $\sim$30 min). The suspension was left for sedimentation (12 hours) and the supernatant filtered through an acetone soluble filter (Millipore WCVP4700). The thickness of the layer was controlled by the amount of the filtered suspension. The
filter was dissolved in acetone and the remaining nanotube thin film
was transferred to a 1cm x 1cm x 1mm quartz substrate (suprasil). At
the end mild heat treatment (50 $^{\circ}$C, few minutes) was used
to remove the solvent from the samples. Transmission spectra vs. wavelength of the thinnest networks of each type and the optical density derived from those are shown in Fig. \ref{fig:Tspectra}.

\begin{figure}[h]
\centering
\includegraphics[width=8cm]{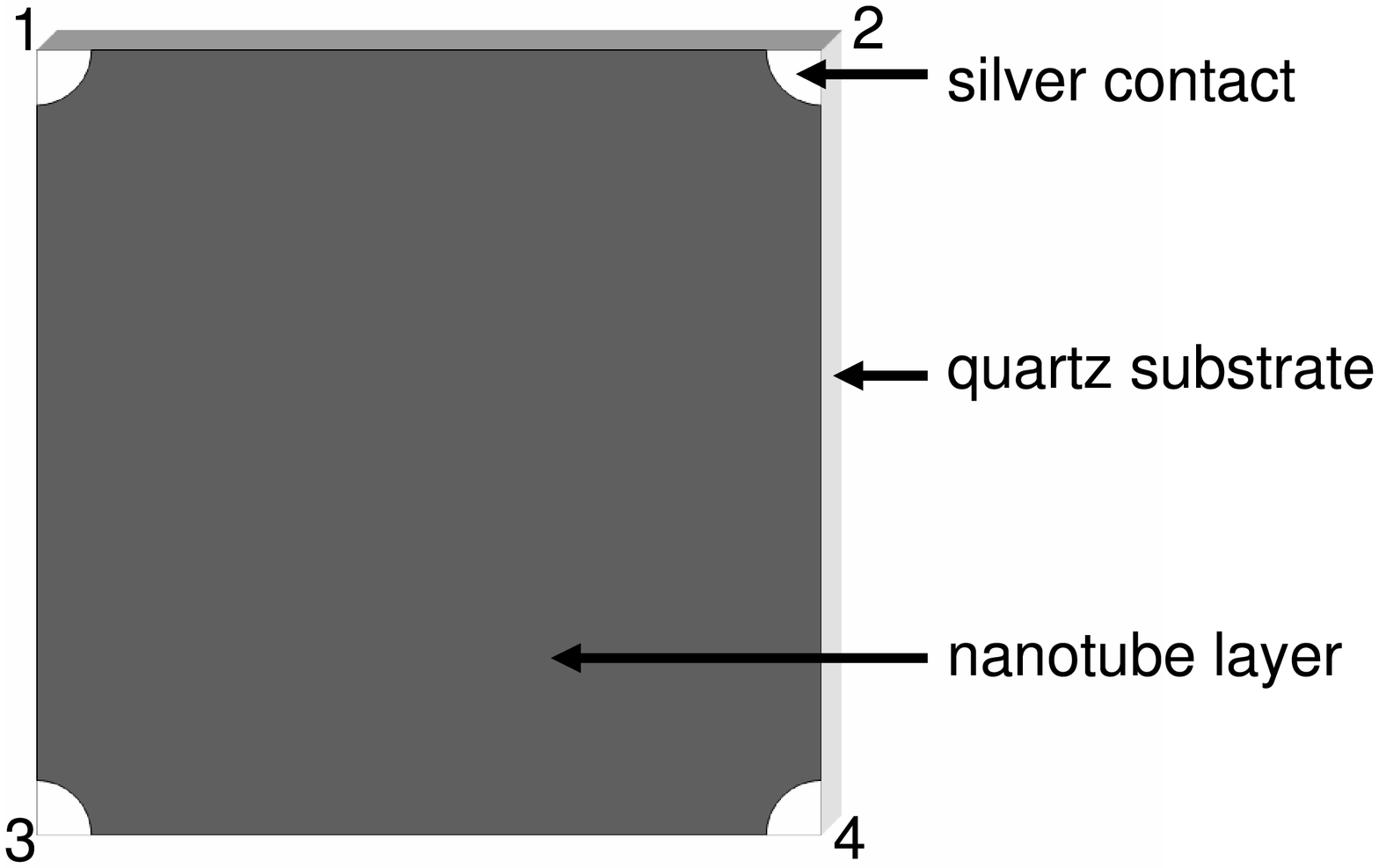}
\caption{}
\label{fig:sample}       
\end{figure}

To measure sheet resistivity, we used the four point van der Pauw method. \cite{vanderpauw58} Figure \ref{fig:sample} shows the measurement
geometry. Contacts were made by painting small dots of colloidal
silver on the four corners of the sample.

Different combinations of voltage and current contacts were used to
improve the accuracy of the resistivity measurement.

\begin{equation}
R_v=\frac{R_{12,34}+R_{34,12}}{2},
\end{equation}
and
\begin{equation}
R_h=\frac{R_{23,41}+R_{41,23}}{2},
\end{equation}
where $R_{ab,cd}$ means we drive the current through the contacts
$a$, $b$ and measure the voltage between $c$, $d$. We measured the
four point resistivity with a Keithley 192 digital multimeter. The
sheet resistivity ($R_{\Box}$) can be calculated using the van der
Pauw formula:
\begin{equation}
e^{-\pi R_v/R_{\Box}}+e^{-\pi R_h/R_{\Box}}=1.
\end{equation}

We measured the near-infrared spectrum of the samples with a Bruker
Tensor 37 FTIR spectrometer and the visible-ultraviolet spectrum
using a Jasco v550 spectrophotometer. Resolution was 2 cm$^{-1}$ in the NIR and 1 nm in the visible/UV.

Figure \ref{fig:ours} shows the sheet conductance vs. optical density curves for nanotube films with varying thickness at two wavelengths indicated in Fig. \ref{fig:Tspectra}: 550 nm and 1600 nm, respectively. In all cases, the measured points can be very well fitted with a linear function. A higher slope of the fitted line indicates a higher figure of merit, since it means higher sheet conductance at a given transmission value. In this respect, it plays a similar role as the $\sigma(0)/\sigma(\omega)$ ratio used in Eqn. \ref{eq:dresselgruner}, and differs from the Haacke figure of merit in an important aspect. The latter gives a number for a film with given optical functions and \emph{a given thickness,} whereas the former two are independent of thickness. The points on the individual curves in Fig. \ref{fig:ours}, however, represent films of increasing thickness as we go farther from the origin. It seems that the morphology of the samples does not influence the figure of merit as in inorganic systems\cite{gordon00} where it was predicted to increase with increasing thickness.

It is remarkable that the curves do not pass through the origin but all show approximately the same intercept at zero sheet conductance. This phenomenon is in agreement with the percolation behavior found by Hu, Hecht and Gr\"uner,\cite{hu04} and the optical density value can be related to the percolation threshold. This finite percolation threshold modifies the meaning of $M(\omega)$ somewhat, since Eqn. \ref{eq:slope} is valid only for lines cutting through the origin. In the case of real nanotubes the equation determining the slope is:
\begin{equation}
\label{eq:slope2} M(\omega) = \frac{S_\Box}{-log\
\frac{T(\omega)}{T_p(\omega)}}
\end{equation}
where $T_p$ is the transmission value at zero sheet conductance.

$M(\omega)$ and $T_p$ can be determined by measuring samples with
different thickness and fitting the data using a linear fit. This means that we need at least two points for a given kind of material. The such obtained slopes M($\lambda$) at two representative wavelengths, 550 and 1600 nm, are summarized in Table \ref{table:slopes}.

\begin{table}
\caption {Figure of merit (M) values at two wavelengths.
\label{table:slopes}}
\begin{tabular}{lcc}
\hline
Sample & 550 nm  & 1600 nm  \\
\hline
arc & 0.0022  & 0.0039  \\
HiPco &  0.0068& 0.0209  \\
CoMoCat CG &  0.0035 & 0.0076 \\
CoMoCat SG & 0.0013 & 0.0032 \\
\hline
\end{tabular}
\end{table}

Both in the visible and near-infrared region, the qualitative order of the networks investigated is identical: the highest figure of merit is found for the HiPco networks, followed by CoMoCat CG, then arc and finally CoMoCat SG.  It is instructive to analyze in detail the two parameters contributing to the M value, as both are equally significant.

The spectra are all relatively flat around 550 nm, thus the transmission values are similar. (As we do not know the thickness of the individual networks, the spectra in Fig. \ref{fig:Tspectra} are not scaled and illustrate only the wavelength dependence.) The order of M values is determined by both the intrinsic conductivity of the nanotubes and the percolation properties of the networks. (This also means that the order found here is only indicative of the specific networks employed in this study and can by no means be generalized to compare different types of carbon nanotube samples.) The low-frequency optical absorption is a good indicator of the intrinsic conductivity; from previous studies\cite{kamaras08} we know that the CoMoCat SG has a significantly lower absorption than the other three, in agreement with the present data.

In one case, we can allow some generalization: this is the trend in the two CoMoCat samples where the Special Grade sample is enriched in semiconducting tubes with respect to the Commercial Grade one. This particular composition exhibits much lower conductivity and therefore sheet conductance than any of the other materials investigated in our study, leading to a smaller M value; this result shows that our method is also useful when the goal is less, and not more, metallic content.

At 1600 nm (right panel of Fig. \ref{fig:ours}) the qualitative order remains the same, but the slopes change. All M values increase but to a different extent. Therefore, the curves for CoMoCat SG and the arc sample almost coincide, because of the increase in $(-log~T)$ in the latter.

In order to add predictive power to the procedure, we indicate
the "optimal" region (above a "threshold" sheet conductance and a "threshold" transmission value) in the upper-left corner of the plot. We used the threshold values $S_\Box$ = 0.007 $\SBox$ and $T = 0.7$ given by Green and Hersam,\cite{green08} but the region can be easily tailored according to specific applications. Thus the desired direction of improvement is straightforward. In this specific case, the HiPco network comes closest to ideal
both at visible (550nm) and infrared (1600nm) wavelengths, but not even this sample will
reach the optimal range by simply thinning the network; instead, fundamental materials properties have to be improved further to increase the slope.

\begin{figure}[h]
\centering

\includegraphics[width=7cm]{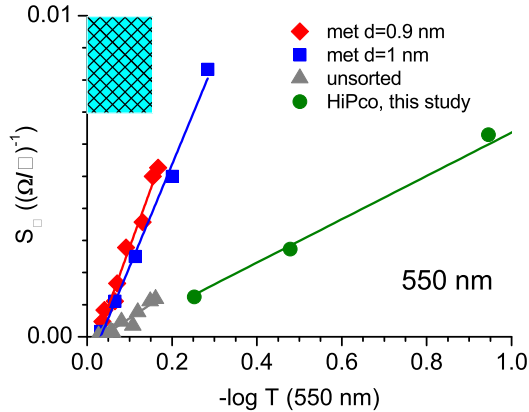}

\caption{The symbols denote dc sheet conductance $S_\Box$ vs. $(-log~T)$ for unsorted and metal-rich HiPCO films (replotted from Ref. \onlinecite{green08}).Lines and shaded area are derived as in Fig. \ref{fig:ours}.  }
\label{fig:merit}       
\end{figure}

One very promising way for such improvement is the already mentioned procedure of separation by density gradient centrifugation. From Eqn. \ref{eq:slope} it follows that enrichment in metallic tubes should increase N$_1$/N$_2$ and thus the M value.

In Fig. \ref{fig:merit} we replot the data of Green and Hersam\cite{green08} at 550 nm for unsorted and metal-enriched nanotube networks in the manner indicated above, along with our unsorted HiPco sample.  The M value for the unsorted sample (0.0088 $\SBox$) is similar to that found in our HiPco networks (0.0068 $\SBox$) and improves significantly in the fractions enriched in metallic tubes (0.032 and 0.037 $\SBox$ for the 1 nm metallic and the 0.9 nm metallic fraction, respectively). It is obvious from the plot that the last one is already very close to the desired range.

Similar analysis can be conducted for transparent conducting oxides like ITO, and also for graphene which shows a qualitatively similar spectrum,\cite{mak08,choi09} but care is to be exercised when applying it to thin metal films since the optical density of those is not necessarily proportional to their thickness. The same is true for carbon nanotube films in the far- and midinfrared region.\cite{pekker06}

\section{Conclusion}

We propose the slope of the sheet conductance $S_\Box = \frac{1}{R_\Box}$ vs. optical density $(-log~T)$ lines as a figure of merit for transparent conductive nanotube coatings for the visible/near IR frequency range. This figure has the advantage of requiring a very simple arithmetic treatment of the data including a linear fit; the resulting graphical representation is easy to handle and makes it simple to predict the direction of optimizing the sample. We demonstrated by comparing single-walled nanotube networks from different sources that the optimal conditions can be reached in other wavelength ranges than the 550 nm typically employed for solar cell applications.

\begin{acknowledgements}

We gratefully acknowledge illuminating discussions with David B. Tanner and Mark C. Hersam. This work was supported by the Hungarian National Research Fund (OTKA) through grants No. 75813 and 67842.

\end{acknowledgements}

\bibliography{PekkerKamarasFigureOfMerit}

\end{document}